\documentclass[10pt,conference]{IEEEtran} 
\IEEEoverridecommandlockouts
\usepackage{cite}
\usepackage{amsmath,amssymb,amsfonts}
\def\BibTeX{{\rm B\kern-.05em{\sc i\kern-.025em b}\kern-.08em
    T\kern-.1667em\lower.7ex\hbox{E}\kern-.125emX}}
    
    \usepackage{algorithm}
\usepackage{algpseudocode}

\usepackage{graphicx}

\usepackage{listings}
\usepackage{caption}
\usepackage{xcolor}
\usepackage{url}

\newcommand{\mysf}{\sf \small \mbox{}}

\lstdefinelanguage{Alloy}{
    keywords = {module,open,abstract,sig,one,extends,run,pred,fun,fact,all,some,alone,Int,let,first,next,nexts,as,disj,lte,plus}
}

\definecolor{codegreen}{rgb}{0,0.6,0}
\definecolor{codegray}{rgb}{0.5,0.5,0.5}
\definecolor{codepurple}{rgb}{0.58,0,0.82}
\definecolor{backcolour}{rgb}{0.95,0.95,0.92}

\lstdefinestyle{mystyle}{
    backgroundcolor=\color{backcolour},   
    commentstyle=\color{codegreen},
    keywordstyle=\color{magenta},
    numberstyle=\tiny\color{codegray},
    stringstyle=\color{codepurple},
    basicstyle=\ttfamily\footnotesize,
    breakatwhitespace=false,         
    breaklines=true,                 
    captionpos=b,                    
    keepspaces=true,                 
    numbers=left,                    
    numbersep=5pt,                  
    showspaces=false,                
    showstringspaces=false,
    showtabs=false,                  
    tabsize=2
}

\begin{document}

\title{Minimizing the Number of Teleportations in Distributed Quantum Computing Using Alloy}



\author{\IEEEauthorblockN{Ali Ebnenasir}
\IEEEauthorblockA{\textit{Department of Computer Science} \\
\textit{Michigan Technological University}\\
Houghton MI 49931, U.S.A. \\
aebnenas@mtu.edu}
\and
\IEEEauthorblockN{Kieran Young}
\IEEEauthorblockA{\textit{Department of Computer Science} \\
\textit{Michigan Technological University}\\
Houghton MI 49931, U.S.A. \\
kdyoung@mtu.edu}
}

\maketitle

\begin{abstract}
This paper presents a novel approach for minimizing the number of teleportations in Distributed Quantum Computing (DQC) using formal methods.  Quantum teleportation plays a major role in communicating quantum information. As such, it is desirable to perform as few teleportations as possible when distributing a quantum algorithm on a network of quantum machines. Contrary to most existing methods which rely on graph-theoretic or heuristic search techniques, we propose a drastically different approach for minimizing the number of teleportations through utilizing formal methods. Specifically, the contributions of this paper include: the formal specification of the teleportation minimization problem in Alloy, the generalizability of the proposed Alloy specifications to quantum circuits with $n$-ary gates,  the reusability of the Alloy specifications for different quantum circuits and networks,  the simplicity of specifying and solving other problems such as load balancing and heterogeneity, and the compositionality of the proposed approach. We also develop a software tool, called qcAlloy, that takes as input the textual description of a quantum circuit, generates the corresponding Alloy model, and finally solves the minimization problem using the Alloy analyzer. We have experimentally evaluated qcAlloy for some of the circuits in the RevLib benchmark with more than 100 qubits and 1200 layers, and have demonstrated that qcAlloy outperforms one of the most efficient existing methods for most benchmark circuits in terms of minimizing the number of teleportations. 
\end{abstract}

\begin{IEEEkeywords}
Distributed Quantum Computing, Teleportation, Formal Methods, Alloy
\end{IEEEkeywords}
\section{Introduction}
\label{sec:intro}

Minimizing the communication costs of distributed Quantum Algorithms (QAs) is an important problem towards realizing large scale quantum computations in Noisy Intermediate-Scale Quantum (NISQ) era where quantum machines are noisy and have limited qubit capacities. One of the major approaches for enabling Distributed Quantum Computing (DQC) on networks of quantum machines includes the mapping of quantum algorithms/circuits to networks, where logical qubits and gates are mapped to  machines towards minimizing quantum communications. While cat-entanglement and quantum teleportation can both enable communication in quantum networks, cat-entanglement is mainly used for sharing read-only qubits in binary gates, which should be disentangled after the communication. Moreover, the cat-entangled qubits  suffer from higher degrees of decoherence. As such, teleportation remains the main mechanism for quantum communication, at the cost of consuming some entangled pairs as resources. Thus, mapping a quantum circuit to a network must be done while minimizing the number of teleportations.

Most existing methods for minimizing the communication costs of DQC utilize a variety of existing heuristics (e.g., genetic algorithms, simulated annealing) and optimization techniques  (e.g., graph partitioning) to tackle the Teleportation Minimization Problem (TMP), which makes them less flexible to change in circuit/network constraints, and less reusable from one circuit/network to another. For example, Andres-Martinez and Heunen \cite{andres2019automated}  reduce the minimization problem to the problem of hypergraph partitioning where   the number of cuts in the partitioned graph must be minimized.  Davarzani  {\it et al.}   \cite{davarzani2020dynamic} create a bipartitie graph out of a quantum circuit where the two sets of vertices include the qubits and the gates. Daei {\it et al.}   \cite{daei2020optimized} model a quantum circuit as a weighted undirected graph, and apply the graph partitioning method of Kernighan–Lin \cite{kernighan1970efficient}. Nikahd  {\it et al.}    \cite{nikahd2021automated} present a window-based partitioning method, and formulate the problem as an instance of Integer Linear Programming (ILP).  While the aforementioned methods provide efficient ways for minimizing the number of teleportations, they suffer from a few important challenges. First, they mostly focus on circuits that have only binary or ternary gates, assuming that circuits with gates of larger arity can be transformed into circuits with binary gates such as CNOT and CZ (controlled phase shift). In theory, it is possible to perform such transformations \cite{kitaev1997quantum}, but it is unclear how efficient they can be done in practice  \cite{dawson2006solovay}. Second, when the  problem constraints change (e.g., network topology, qubit capacity of nodes), the entire problem formulation should be regenerated (e.g., recreate a new  hypergrpah), and there is little room for reusing the problem specification. Third, most existing problem formulations (e.g., graph-theoretic, ILP) are difficult to understand by mainstream designers, thereby making it hard to fine-tune them when problem parameters change or computational costs become prohibitive.

\begin{figure}
\centering
\includegraphics[scale=0.1]{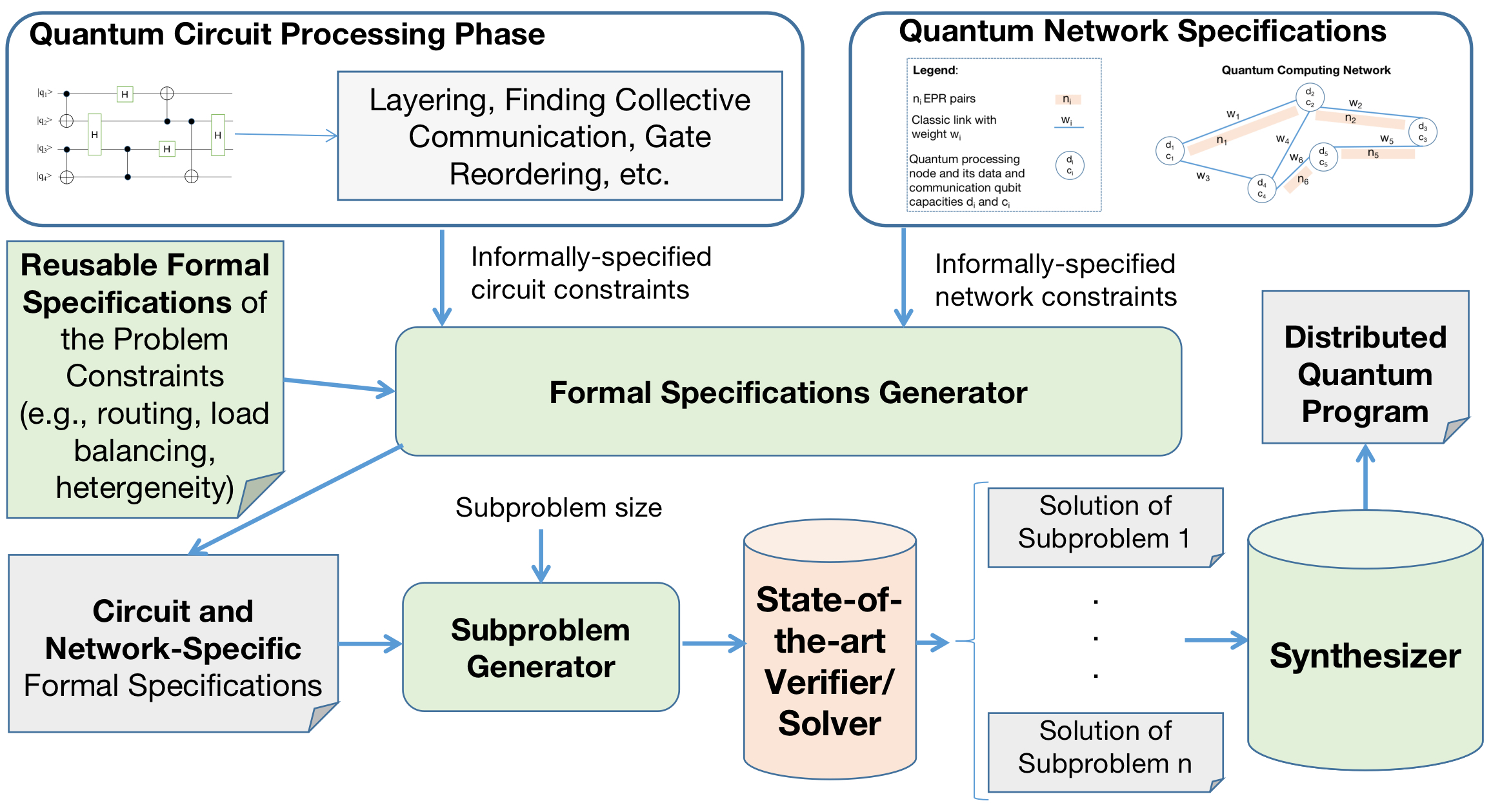}
\caption{A vision for the mapping of QAs to quantum networks.} 
\label{fig:vision}
\end{figure}

This paper presents a novel vision of using formal methods for high-level and reusable specification of the TMP, where part of the formal specifications of the problem constraints remain unchanged and can be reused even if the circuit or network characteristics change. Specifically, our vision for mapping QAs to networks  is heavily reliant on formal specification languages and compositional problem solving (see Figure \ref{fig:vision}).   Such formal specifications are at a high level of abstraction and easily understandable for mainstream  engineers (compared with ILP or hypergraph representations).   The generator of formal specifications in Figure \ref{fig:vision} takes in the reusable formal constraints of TMP, and then generates a circuit and network-specific specification. Subsequently, the generated specification is divided into a set of subproblems. Each subproblem is then solved independently using state-of-the-art verifiers/solvers, and their solutions are combined using a synthesizer.

The proposed approach in this paper is an instantiation of the framework in Figure \ref{fig:vision} for the Alloy specification language \cite{jackson2012software}, where the Alloy specification of TMP includes a circuit and network-independent part, which is reusable, and a circuit and network-dependent section (a.k.a., problem-specific section). The reusable part of the Alloy specifications of TMP includes the relations that govern the movement of qubits,  their allocation on machines, and how these relations change from one circuit layer to another, where a  {\it layer} includes a set of gates that  can be executed concurrently on a disjoint set of qubits. The problem-dependent part of the Alloy specifications formally describe the circuit structure as well as network constraints (e.g., topology, communication costs, qubit capacity of machines). 
The basic units of abstraction in Alloy specifications (a.k.a. models) include {\it signatures} and {\it relations}, where a signature captures a type/set, and relations establish ties between signatures. For example, in an Alloy model, a qubit $q$ can be declared as a signature, and then associated with another signature of type `Machine' where $q$ is located. Alloy enables the navigation of relations using the `dot join' operator borrowed from relational algebra. Constraints in Alloy are specified as first-order logic statements, and can either be enforced on an Alloy model, or queried.   The combination of the two parts provides us an Alloy model that is ready for verification by the Alloy analyzer, which contains state-of-the-art SAT solvers. The analyzer verifies whether the number of teleportations performed  is at most equal to an upper bound. Selecting a sufficiently small  upper bound will force the analyzer to find a minimum solution. 
Depending on the scale of the quantum circuit, we may divide the problem into several subproblems, each focusing on a subcircuit containing some number of layers.

We have implemented the proposed method as a software tool, called qcAlloy, which is available at \url{https://github.com/KieranYoung/Alloy-DQC}. The results of our experimental evaluations on the RevLib benchmark \cite{revLib2005,wille2008revlib} indicate that qcAlloy outperforms one of the most efficient methods \cite{nikahd2021automated}  in terms of the number of teleportations in most cases up to 50\% while underperforming in a few cases (e.g., QFT circuit). qcAlloy efficiently solves TMP for circuits with more than 100 qubits and 1200 layers. 

\noindent{\bf Organization}.\  Section \ref{sec:prelim} represents some basic concepts of quantum circuits and Alloy. Section \ref{sec:prob} states the TMP problem. Section \ref{sec:alloySpec}  specifies the constraints of  TMP in Alloy. Section \ref{sec:loadb}  investigates the TMP problem in the context of load balancing, and Section \ref{sec:hetero} studies TMP in heterogeneous networks. Subsequently, Section \ref{sec:eval} presents our experimental results.  Section \ref{sec:rel} discusses related work. Finally, Section \ref{sec:concl} makes concluding remarks and discusses future work. 

\section{Preliminaries}
\label{sec:prelim}
This section presents some basic concepts of quantum circuits and their abstract representation in 
Subsection \ref{sec:circ}. 
 Subsection \ref{sec:alloy} provides an overview of the Alloy language \cite{jackson2012software} and its analyzer.

\subsection{Quantum Circuits and Circuit Graphs}
\label{sec:circ}
\begin{figure}
\centering
\includegraphics[scale=0.12]{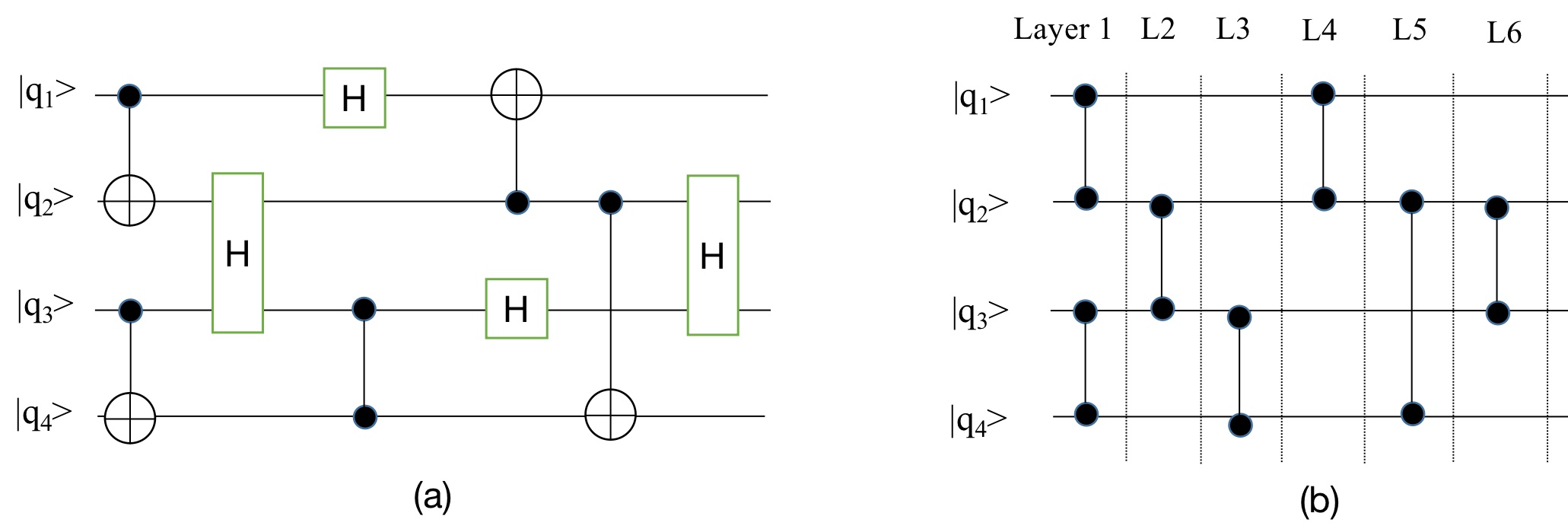}
\caption{Abstracting circuit (a) as the circuit graph (b).} 
\label{fig:abscirc}
\end{figure}

Quantum Algorithms (QAs) capture the logic and order of quantum transformations that are performed on quantum information bits (i.e., {\it qubits}) towards solving a problem. A common approach for representing QAs includes quantum circuits that contain a set of horizontal wires carrying quantum information from left to right and quantum gates applied on a subset of wires vertically \cite{nielsen2001quantum}. There is a one-to-one correspondence between the input qubits of a circuit and its wires. 
For example, Figure \ref{fig:abscirc}-(a) illustrates a quantum circuit processing four qubits through applying a set of gates. A quantum machine runs a circuit by executing its gates from left to right.  Notice that, some gates have a single qubit as their input and some take two qubits (a.k.a. {\it binary gates}) in Figure \ref{fig:abscirc}-(a). In general, gates might have multiple input qubits; however, it is known  \cite{kitaev1997quantum} that any quantum circuit can be represented by another circuit formed of a universal set of single-qubit and binary gates to some degree of accuracy. In this paper, we consider an abstract representation of quantum circuits, called the {\it circuit graph} (e.g., see Figure\ref{fig:abscirc}-(b)), where we consider a vertex as a point of intersection between a wire and a gate, and an edge represents a binary gate. Multi-qubit gates are captured as multiple connected edges (i.e., path).

Since the Noisy Intermediate Scale Quantum (NISQ) machines have a limited qubit capacity (due to sensitivity to environmental noise and quantum decoherence), large quantum circuits cannot be executed locally on a single machine and they have to be partitioned and distributed over a network of quantum machines. Circuit partitioning is performed horizontally where proper subsets of qubits are assigned to different machines in the network. As a result, the inputs of some gates may be located in different machines, called {\it global gates}. One way to execute a global gate $g$ is to move all its remote input qubits to the machine that is supposed to execute $g$, thereby turning $g$ into a {\it local gate}. 

Circuit partitioning requires a means for the transmission of quantum information from one machine to another. However, quantum information cannot be copied \cite{wootters1982single}, nor can it be communicated without error. A reliable way of communicating quantum information includes {\it teleportation} 
\cite{bennett1993teleporting}, which is a method for transferring quantum information from one place to another. To teleport a qubit of information from a location $loc_0$ to another location $loc_1$, we need two classic bits of information as well as an Entangled Pair (EP) of qubits that are already distributed over $loc_0$ and $loc_1$. The teleportation of a qubit from $loc_0$ to $loc_1$ can then be achieved through the execution of some local quantum gates in $loc_0$ and $loc_1$. Thus, the communication cost of a Distributed Quantum Circuit (DQC) boils down to the total number of teleportations performed. As such, we would like to partition a circuit in such a way that requires minimum number of teleportations.

\subsection{Alloy and Scope-Based Model Checking}
\label{sec:alloy}

Alloy \cite{jackson2012software} is a declarative specification language, which combines first-order logic with relational algebra towards providing  a highly expressive specification language. Alloy has been used for solving a variety of problems in different domains of applications (\url{https://alloytools.org/citations/case-studies.html}), which demonstrate its expressiveness and its practical use. The basic unit of abstraction in Alloy is a relation, which can be of an arbitrary arity (e.g.,  unary, binary, ternary). To define relations over abstract entities, one needs to specify such entities first. An abstract entity (i.e., a unary relation) in Alloy is captured as a {\mysf Signature}, which can be considered as a set of elements, called {\it atoms}. For example, to model a directed graph, one can declare a signature for graph nodes as follows: {\mysf sig Node\{\}}, where  {\mysf sig} is a keyword for declaring signatures in Alloy. To specify the arcs of a directed graph, we can define a binary relation {\mysf arc} that starts at a node and ends at another node:  {\mysf sig Node\{  arc: one Node \}}. From an object-orientation point of view, one can think of the signature {\mysf Node} as a class and the relation {\mysf arc} as a field in that class. The domain of the {\mysf arc} relation includes the atoms of {\mysf Node} and its range is {\mysf Node} too. Alternatively, {\mysf arc}  can  be  specified as `{\mysf  Node $\rightarrow$ one Node}', where  $\rightarrow$ denotes the {\it cross product} operation applied on two signatures. 
 The quantifier {\mysf one} in `{\mysf one Node}' stipulates that each atom of signature {\mysf Node}  is mapped to exactly one atom in signature {\mysf Node} under the relation {\mysf arc}.

In principle, the cardinality of {\mysf Node} could be infinite, however, to enable automated analysis, Alloy requires us to specify an upper bound, called the {\it scope} of {\mysf Node}, on the size of {\mysf Node} when it comes to verifying any sort of constraint about {\mysf Node}. For example, if we were to check whether there are cycles in a graph, then we would specify an assertion as follows: `{\mysf assert isCyclic\{ some n:Node $\mid$ n in n.\string^arc\}}'. Intuitively, the assertion stipulates that there is some node $n$ that is in its dot join product with the  transitive closure, denoted \string^, of the relation {\mysf arc}. In other words, we are asking whether $n$ can be reached from itself through one or more steps in the graph. This is achieved through the dot join operator `.', which is similar to relational join in databases. To check the assertion using the Alloy analyzer, we write the following command: `{\mysf check isCyclic for 5 Node}'.  In this case, an {\it instance} of our Alloy model includes a graph with $d$ nodes, where $1 \leq d \leq 5$, and some arcs inserted between the nodes in a non-deterministic fashion. The Alloy analyzer creates all instances that adhere to the constraints of the Alloy specification, and then checks the  {\mysf isCyclic} assertion on each instance. Each instance is a graph with up to 5 nodes because we specified the scope as `{\mysf for 5 Node}' in the `{\mysf check isCyclic}' command. The default scope is three in case no scope is explicitly given to the analyzer. 

Alloy specifications may include constraints that must hold across all  instances of the specification, called {\it facts} (a.k.a {\it invariant} constraints). For instance, in our graph example, if we are interested in instances that have  no self-loops, then we add the following fact to our alloy specification: `{\mysf fact noSelfLoop \{ no n:Node $\mid$ n in n.arc\}}', where {\mysf n.arc} returns the set of nodes to which  the node $n$ is connected, and `in' is the keywork for subset operator in Alloy. ({\mysf  noSelfLoop}  is an optional user-defined identifier.) One can define {\it functions} in Alloy, where the parameters of a function are relations and its return value is some relation too. For example, to return the nodes from where there is an arc to a specific node $n$, we  write the following function: `{\mysf fun incomings[n: Node]:Node\{ arc.n \}}'. The dot join of {\mysf arc} and $n$ returns a set of nodes whose every member $n'$ is the first element of some ordered pair $(n', n)$ in relation {\mysf arc}. A special kind of function includes a {\it predicate}, which returns only Boolean values. For instance, the predicate `{\mysf pred isArc[n1, n2:Node]\{ (n1 $\rightarrow$ n2) in arc\}}' returns true iff (if and only if) the {\mysf arc} relation includes an ordered pair $(n1,n2)$ between two nodes $n1$ and $n2$; i.e., an arc from $n1$ to $n2$ exists in the graph. To check the validity of a {\mysf isArc}, we run it as follows: `{\mysf run isArc}'. The Alloy analyzer searches through all possible instances of the model, in this case directed graphs  each having up to 3 nodes (because we have not explicitly specified a scope for the {\mysf run} command). Then, the analyzer returns all those instances that adhere to the constraints of the Alloy specification (e.g., facts) as well as the constraints of the predicate/function.

\section{Problem Statement}
\label{sec:prob}

This section states the Teleportation Minimization Problem (TMP) as follows:

\begin{itemize}
\item {\it Input}: A circuit graph $G$ with $L$ layers on $n > 1$ qubits,  a function $f: Q \mapsto M$ that captures the initial allocation of a set $Q=\{q_1, \cdots , q_n\}$ of $n$ qubits to a set $M=\{m_1, \cdots , m_k\}$ of $k$ quantum machines, where $k < n$, and a set $C=\{c_1, \cdots , c_k\}$, where $c_i$ denotes the capacity of $m_i$, for $1 \leq i \leq k$. A special case is when all $c_i$ values are equal; i.e., a {\it homogeneous} network.

\item {\it Output}: The minimum number of teleportations that must be performed to ensure that all  gates can be executed locally. The output should also include $L$ functions $f_1, f_2, \cdots , f_L : Q \mapsto M$ representing how teleportations are performed from one layer to another. 
\end{itemize}

A restricted statement of the problem may require that there is no machine that always  remains vacant while solving the problem. That is, the function $f$ must be onto in almost every layer of the circuit. This problem relates with load balancing and has conflicting requirements with minimizing teleportations, which we shall discuss in Section \ref{sec:loadb}.

\section{Specifying the TMP in Alloy}
\label{sec:alloySpec}

This section presents a novel approach for specifying the  teleportation minimization problem in Alloy. Section \ref{sec:binary} specifies a special case of the problem for quantum circuits with binary gates, and Section \ref{sec:nary} generalizes the proposed approach to circuits with $n$-ary gates, for $n>2$. We discuss the reusability of Alloy specifications in Section \ref{sec:reuse}.

\subsection{Circuits With Binary Gates}
\label{sec:binary}

\begin{figure}
\centering
\includegraphics[scale=0.14]{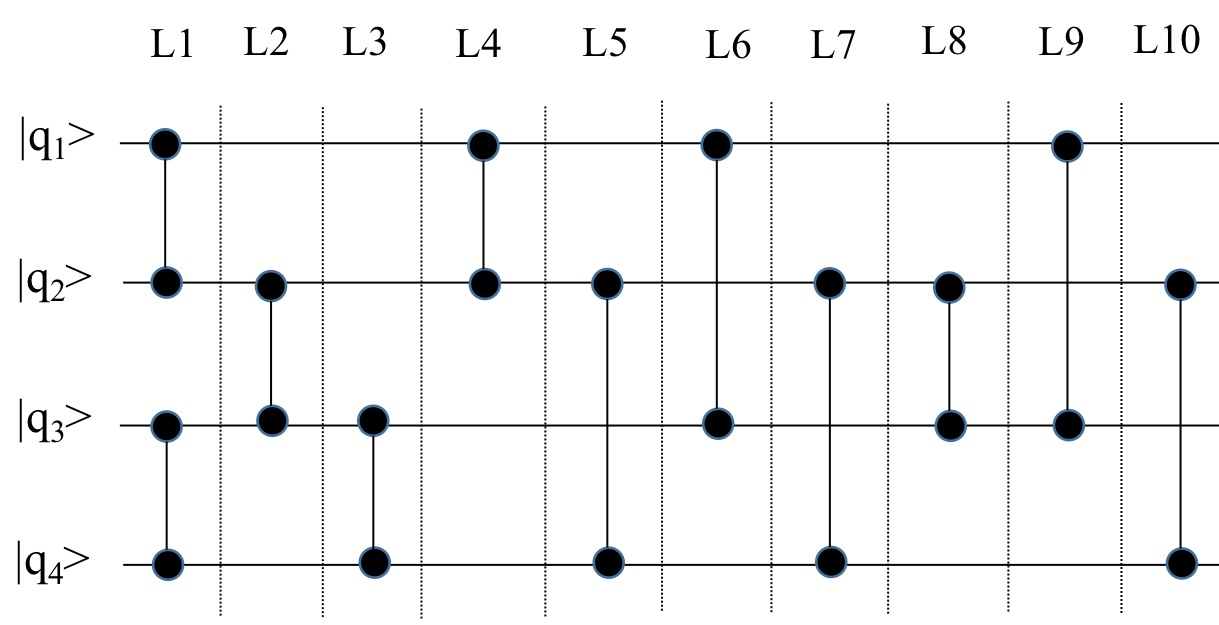}
\caption{Circuit 1 with binary gates and depth 10.} 
\label{fig:circ1}
\end{figure}

This section investigates a special case of the problem for quantum circuits that have only single-qubit and binary gates. For example, the circuit graph in Figure \ref{fig:circ1} has 4 qubits and 10 layers. To formulate the problem in Alloy, we first design abstractions that model a circuit graph. Each circuit graph has some qubits and some binary gates, abstracted as edges in the circuit graph. For this reason, we consider an abstract signature {\mysf Qubit} and extend four singleton qubits from it (in Lines 5 and 6 of Listing \ref{lst:one}). The semantics of an {\it extension} is similar to a subtype. That is, the singleton signatures $q1, q2, q3, q4$ (denoted by `{\mysf one sig}') are of type {\mysf Qubit}. The {\mysf abstract} modifier indicates that {\mysf Qubit} is partitioned by its subtypes.
We also specify the notion of a quantum machine as an abstract signature and extend two singleton signatures $M1$ and $M2$ to represent a network of two machines. The number of machines also represents the number of subgraphs to which the circuit graph is partitioned in a graph-theoretic sense. As the circuit executes, some qubits might be teleported from some machines to other machines (depending on network topology) to ensure that all binary gates in the next layer are local.

\noindent{\bf Modeling the circuit graph and its layers}.\ To model the circuit graph and its transitions from one layer to another, we consider the signature {\mysf circGraph} (Lines 11 to 18 in Listing  \ref{lst:one}). Each atom of {\mysf circGraph} models a layer of the circuit  through the  {\mysf edges} relation (Line 12), the  {\mysf location} relation (Line 14) and the  {\mysf numTele} relation (Line 16). 
To capture the total ordering of circuit layers from left to right, we impose a total ordering on all atoms of {\mysf circGraph}  by specifying `{\mysf open util/ordering[circGraph] as grph}' in Line 2 of Listing  \ref{lst:one}. For example, if in an instance of our Alloy model, {\mysf circGraph} has 10 atoms, the Alloy analyzer searches through all possible total orderings of these atoms for a solution. We  formalize such orderings as $c_0, c_1, \cdots, c_{9}$, where each $c_i$ is an atom of {\mysf circGraph}. We  refer to such orderings by the identifier `grph' in our Alloy specification. 
The {\mysf edges} relation captures the edges of the circuit graph. These edges show which gates need to be local, and where the corresponding input qubits should be. The allocation of qubits to machines in each circuit layer is captured by the {\mysf location} relation. Notice that,   {\mysf edges} is a ternary relation with the cross product {\mysf circGrpah} $\rightarrow$ {\mysf Qubit} $\rightarrow$ {\mysf Qubit}, however, its domain is implicit and includes the signature in which it is declared (i.e., the {\mysf circGraph} signature). Same holds for the {\mysf location}  relation, which represents the mapping of qubits to machines; i.e., the ternary relation  {\mysf circGrpah} $\rightarrow$ {\mysf Qubit} $\rightarrow$ {\mysf Machine}. In other words,  each atom of  {\mysf circGrpah} is associated with two binary relations {\mysf Qubit} $\rightarrow$ {\mysf Qubit} and {\mysf Qubit} $\rightarrow$ {\mysf Machine}, thereby creating ternary relations. The binary relation  {\mysf numTele} (specified as   {\mysf circGrpah} $\rightarrow$ {\mysf Int}) associates an integer to each atom $c_i$ of {\mysf circGrpah} capturing the total number of teleportations performed up to $c_i$ in the total ordering $grph$.

\begin{lstlisting}[caption= Alloy code for circuit graph of Figure \ref{fig:circ1},basicstyle=\small, language=Alloy,keywordstyle={\bf },label=lst:one]
module teleport
open util/ordering[circGraph] as grph
open util/integer

abstract sig Qubit {  }
one sig q1,q2,q3,q4 extends Qubit{} 

abstract sig Machine { } 
one sig M1,M2 extends Machine{}

sig circGraph{
  edges:Qubit->Qubit, 
    // Represents the non-local gates 
  location:Qubit->Machine, 
    // Allocates qubits to machines
  numTele:Int 
    // Keeps the number of teleportations.
} 
// Each qubit must be on exactly one machine.
fact qubitAlloc {  
  all q:Qubit,c:circGraph|#c.location[q] =1}
// Capacity of each machine is at most 3 qubits.
fact mCap {  
 all c:circGraph,m:Machine| #(c.location).m < 4}

// Start with qubits q1, q2 on Machine 1 and 
//  q3, q4 on Machine 2.
fact  CircuitGraph {
  let c0=grph/first|
  c0.edges=(q1->q2)+(q3->q4)&&(c0.numTele=0) &&
  c0.location=(q1->M1)+(q2->M1)+ 
	       (q3->M2)+(q4->M2) &&
  let c1=c0.next|c1.edges=(q2->q3) &&
  let c2=c1.next|c2.edges=(q3->q4) &&
  let c3=c2.next|c3.edges=(q1->q2) && 
  let c4=c3.next|c4.edges=(q2->q4) && 
  let c5=c4.next|c5.edges=(q1->q3) && 
  let c6=c5.next|c6.edges=(q2->q4) && 
  let c7=c6.next|c7.edges=(q2->q3) && 
  let c8=c7.next|c8.edges=(q1->q3) && 
  let c9=c8.next|c9.edges=(q2->q4) }

pred teleport[loc: Qubit -> Machine, 
  r:Qubit->Qubit,uloc:Qubit->Machine, 
  tele:Int,utele:Int]. {
// Pairs of qubits related under r must move 
// to the same machine in the next layer.
 all disj q0,q1:Qubit| 
       ((q0->q1 in r)) implies q0.uloc=q1.uloc
// utele contains the number of qubits moved.
	utele=plus[tele,#(uloc-loc)]}

fact layerTransition {
// These are the constraints that rule every 
//   two consecutive atoms of circGraph.
// Think of this as the transitions in a 
//  linear computation.
  all c:circGraph,uc:grph/next[c] { 
  teleport[c.location,uc.edges, 
        uc.location,c.numTele,uc.numTele] }}

// We would like to have at most 6 qubits  
// teleported at the final atom of 
//  the ordering grph.
pred finalLayer {lte[grph/last.numTele, 6] }

// Run this model for a scope of 10 atoms of 
//  type circGraph and integers in the 
//  range 0 to 31.
run finalLayer for 10 circGraph, 5 Int
\end{lstlisting}

The `{\mysf fact}' constraint in Lines 28-41 models the edges of the circuit graph and initializes the {\mysf location} and {\mysf numTele} relations. Notice that, in Layer 1 of Figure \ref{fig:circ1} qubits 1 and 2 form the inputs to a binary gate and the qubits 3 and 4 are inputs to another binary gates. Line 30  specifies {\mysf  c0.edges = (q1 $\rightarrow$ q2) + (q3 $\rightarrow$ q4)} as the edges of the first {\mysf circGraph} atom (i.e., {\mysf c0 = grph/first}) in a total ordering. Initially, we set the number of teleportations to 0 in the first circuit layer; i.e.,  {\mysf c0.numTele = 0}, and the qubit mapping is specified as {\mysf c0.location = (q1 $\rightarrow$ M1) + (q2 $\rightarrow$ M1) + (q3 $\rightarrow$ M2) +(q4 $\rightarrow$ M2)}; i.e., qubits 1 and 2 are located in machine $M1$ and machine $M2$ holds qubits 3 and 4.  Lines 33 to 41 of Listing \ref{lst:one} specify the structure of subsequent layers of the circuit in a total ordering. The `{\mysf next}' keyword is a binary relation defined in the ordering `{\mysf grph}' that points to the next atom of the current atom. Notice that, these lines specify only the {\mysf edges} relation because the relations {\mysf location} and {\mysf numTele} should be determined by the Alloy analyzer. In other words, the solution of the minimization problem is provided in terms of the  {\mysf location} relation and {\mysf numTele}  holds the number of teleportations performed in the corresponding solution. 

\noindent{\bf Invariant constraints}.\ There are some constraints that must hold across all instances of our Alloy model specified as facts. For example, the {\mysf qubitAlloc} fact in Lines 20-21 of Listing \ref{lst:one} specifies that each qubit must be associated with {\em exactly} one machine in each circuit layer. That is, no qubit can remain unallocated, nor can it be allocated to multiple machines. The `{\mysf all}' keyword denotes the universal quantification, and `{\mysf some}' represents the existential quantification. 
The `{\mysf mCap}' fact in Lines 23-24 stipulates that each machine has a qubit capacity of three. The {\mysf layerTransition} fact (in Lines 53-60) imposes some constraints on every consecutive pair of {\mysf circGraph} atoms, formally specified as `{\mysf all c: circGraph, uc: grph/next[c]}', where $c$ and $uc$ denote two consecutive layers of the circuit under the total ordering {\mysf grph}. The predicate {\mysf teleport} (in Lines 43-51) specifies the nature of the constraint imposed on pairs of {\mysf circGraph} atoms, indicating what should occur when moving from one layer to the next layer of the circuit. Specifically, given the current layer $c$, we require that (see Line 48) all distinct  qubits $q0, q1$ (captured by the `{\mysf disj}' keyword) connected by the {\mysf edges} relation of the next layer, denoted by {\mysf uc.edges}, must be in the same machine when we move to the next layer (specified as {\mysf q0.uloc = q1.uloc}). The Alloy analyzer identifies the machine that is the best choice for executing a global gate because the analyzer considers all feasible orderings of the atoms of {\mysf circGraph}. In fact, the Alloy specification of TMP looks at the problem in a global way in contrast to some related work \cite{nikahd2021automated,sundaram2023distributing}  that solve the problem in a greedy fashion.   
Moreover, the  {\mysf teleport} predicate keeps track of the number of moved qubits. Formally, Line 51 states that the number of teleported qubits in the next layer (i.e., {\mysf utele}) equals to the summation of the current number of teleported qubits (i.e., {\mysf tele}) and the number of qubits that were moved in the last transition from the current layer to the next one (i.e., {\mysf \#(uloc - loc)}). The constraint {\mysf \#(uloc - loc)} returns the number of tuples that exist in the updated {\mysf location} relation {\mysf uloc}) which do not exist in the  {\mysf location} relation in the current layer (i.e., {\mysf loc}). The {\mysf plus} function in Line 51 is imported from the `integer.als' file in Line 3, and returns the summation of its two parameters.

\noindent{\bf Minimization predicate}.\ In order to force the Alloy analyzer to find a total ordering with minimum number of teleportations, we query it for orderings whose last {\mysf circGraph} atom has the desired upper bound on the total number of teleportations. Formally, we specify this as the predicate {\mysf finalLayer} in Line 65 where we state that `{\mysf grph/last.numTele}' (i.e., the `{\mysf numTele}' relation in the last atom of the ordering `{\mysf grph}')  must be less than or equal to six. The {\mysf lte} predicate is taken from the `integer' library imported in Line 3. To find a solution, we ask the analyzer (in Line 70) to look for instances that have up to ten atoms in the {\mysf circGraph}  signature and for integers specifiable in five bits. One can execute the model with a very small upper bound (i.e., second parameter of predicate {\mysf lte} in Line 65), and incrementally increase this parameter until a solution is found. Such a solution will have the minimum number of teleportations. 

\begin{figure}
\centering
\includegraphics[scale=0.11]{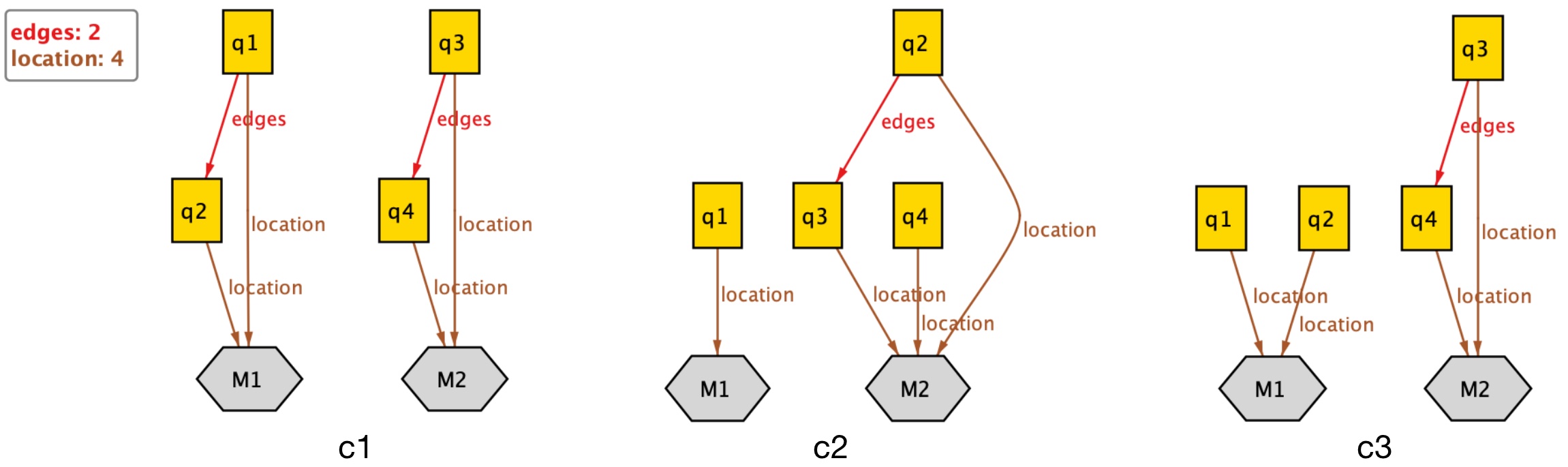}
\caption{Three consecutive layers of a solution generated by Alloy. Arrows represent the binary relations.} 
\label{fig:out1}
\end{figure}

\noindent{\bf Solution visualization}.\ Executing Line 70 of Listing  \ref{lst:one} will search for a total ordering on the atoms of  {\mysf circGraph} where the total number of teleportations is minimized globally. The Alloy analyzer visualizes such an instance, and enables us to go through the circuit layers and see how qubits are teleported from one layer to another. Figure \ref{fig:out1} illustrates the state of  three consecutive layers where $q2$ is teleported to machine $M2$ in the $c2$ layer, and then moved back to machine $M1$ in layer $c3$. When we juxtapose this output with Figure \ref{fig:circ1}, we realize that moving $q2$ to $M2$ is done since in layer $c2$ there is a global binary gate whose inputs include $q2$ and $q3$. Such a teleportation will transform this gate into a local gate that can be execute on $M2$. Table \ref{tab:sol1} demonstrates the teleportations performed in each layer of the circuit of Figure \ref{fig:circ1}. The minimum number of teleportations for this circuit is actually five, which matches with what reported in \cite{sundaram2023distributing}. (Following related work \cite{sundaram2023distributing}, we consider a swap as one teleportation.)

\begin{table}
\begin{center}
\begin{tabular}{||c | c | c | c||} 
 \hline
 Circuit layer & \# of teleported qubits & Machine 1 & Machine 2 \\ [0.5ex] 
 \hline\hline
 1 & 0 & $q1, q2$ & $q3, q4$ \\ 
 \hline
 2 & 1 & $q1$& $q2, q3, q4$ \\
 \hline
 3 & 1 & $q1, q2$  & $q3, q4$ \\
 \hline
4 & 0 & $q1, q2$  & $q3, q4$ \\
 \hline
 5 & 1 (swap) & $q2, q4$  & $q1, q3$ \\
 \hline
 6 & 0 & $q2, q4$  & $q1, q3$  \\ 
 \hline
 7 & 0 & $q2, q4$  & $q1, q3$  \\ 
 \hline
 8 & 1 & $q4$& $q1, q2, q3$ \\
 \hline
  9 & 0 & $q4$& $q1, q2, q3$ \\
 \hline
  10& 1 & $q2, q4$  & $q1, q3$  \\ 
 \hline
\end{tabular}
\end{center}
\caption{A solution for the circuit in Figure \ref{fig:circ1}}
\label{tab:sol1}
\end{table}

\subsection{Circuits With $n$-ary Gates}
\label{sec:nary}

This section extends the results of the previous section to circuits with $n$-ary gates. To the best of our knowledge, this is an open problem as most existing methods \cite{zomorodi2018optimizing} assume binary/ternary gates.  We present a solution for the TMP problem  in a distributed quantum circuit made of arbitrary $n$-ary gates, for a some fixed $n>2$. Reusing the contributions of the previous section, we simply extend our Alloy specification to cases where the circuit may contain gates with $n >2$ input qubits. First, we observe that, the use of $n$-ary gates increases the chances of having global gates in a circuit. This potentially exacerbates the minimization problem because more qubits must be teleported  to ensure that a global gate can be executed locally. Second, the capacity of machines must be at least equal to $n$ because if a global $n$-ary gate is to be executed locally, then the machines must be able to hold at least $n$ qubits.

\begin{figure}
\centering
\includegraphics[scale=0.1]{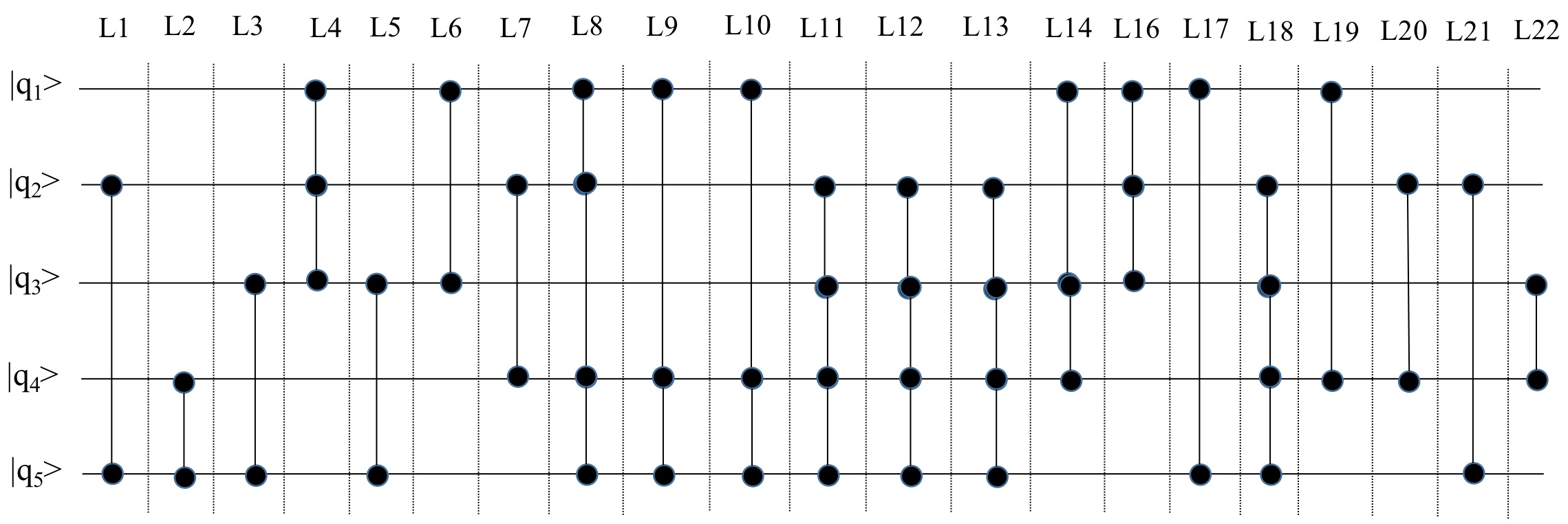}
\caption{Circuit 2 with 4-ary gates and depth 22.} 
\label{fig:circ2}
\end{figure}

The core of the problem in $n$-ary global gates is that all $n$ input qubits must be present in a quantum machine. Such a constraint can easily be captured in our Alloy specification by looking at the input qubits in the circuit graph as a set of edges between the vertices in the layer where the $n$-ary gate is located. These edges obviously intersect at common vertices. For example, the circuit graph in Figure \ref{fig:circ2} (taken from \url{https://reversiblebenchmarks.github.io/hwb5d3.html}) includes three 4-ary gates in layers 11 to 13, each one represented as three connected edges. This circuit has binary, 3-ary and 4-ary gates located in 22 layers over 5 qubits. Lines 26 to 55 of Listing \ref{lst:two} specify the structure of the circuit graph. For example, the 4-ary gate in Layer 11 of Figure \ref{fig:circ2} takes qubits 2-5 as its inputs, and we specify this requirement in Lines 39-40 of Listing \ref{lst:two}  as three pairs {\mysf  (q2 $\rightarrow$ q3)+ (q3 $\rightarrow$ q4)+(q4 $\rightarrow$ q5)}. The common qubits q3 and q4 enforce the constraint that all four qubits must be on the same machine.  We consider a fully connected topology for a network of three machines. Line 70 of Listing \ref{lst:two} runs the Alloy analyzer for 22 instances of the {\mysf circGraph} as the circuit has 22 layers. Table \ref{tab:sol2} illustrates a solution with 13 teleportations. If we consider each swap operation as one  teleportation, then the total number of teleportations will be 11.

\subsection{Reusability of the Alloy Model}
\label{sec:reuse}

A significant advantage of our approach lies in the reusability of a major portion of our Alloy specification which is circuit and network-independent. Specifically, if we consider the Alloy code in Listings  \ref{lst:one} and  \ref{lst:two}, we observe that the only problem-specific  lines of code includes the parts where we specify (1) the number of qubits (e.g., Line 6 in Listing  \ref{lst:two}); (2) the number of quantum machines or circuit partitions (Line 9 in Listing  \ref{lst:two}); (3) the machine capacity on Line 20; (4) the specification of the circuit graph and initial qubit allocation (Lines 23-55 in Listing  \ref{lst:two}); (5) the optimization parameter of the {\mysf finalLayer} predicate (i.e., second parameter of the {\mysf lte} predicate in Line 67 of Listing \ref{lst:two}), and (6) the command for executing the Alloy model (e.g., Line 70 in Listing \ref{lst:two}). For any given quantum circuit/network, one has to specify the aforementioned circuit/network-specific lines of code. In fact, qcAlloy automates the compilation of a given quantum circuit to these circuit-dependent part of the Alloy specification. The circuit/network-independent part of our Alloy specification includes the  signatures, the {\mysf qubitAlloc} and {\mysf layerTransition}  facts, and the {\mysf teleport} and {\mysf finalLayer} predicates. These parts capture the main characteristics/constraints of any solution of the TMP problem and designers need not revise them when circuit/network changes. Next, we shall show how formal specifications and their reusability simplify the task of solving other variants of the TMP under additional constraints.

\begin{lstlisting}[caption= Alloy code for circuit graph of Figure \ref{fig:circ2}, basicstyle=\small, language=Alloy,keywordstyle={\bf },label=lst:two]
module teleport
open util/ordering[circGraph] as grph
open util/integer

abstract sig Qubit { }
one sig q1,q2,q3,q4,q5 extends Qubit{}

abstract sig Machine { } 
one sig M1,M2,M3 extends Machine{}

sig circGraph{
  edges:Qubit->Qubit, 
  location:Qubit->Machine, 
  numTele:Int }
 // Each qubit must be on exactly one machine.
fact qubitAlloc { 
  all q:Qubit,c:circGraph|#c.location[q]=1}
// The capacity of each machine is 4 qubits. 
fact mCap {
  all c:circGraph,m:Machine|#(c.location).m<5} 
fact  CircuitGraph { 
  let c0 = grph/first| 
  c0.edges=(q2->q5)&&(c0.numTele=0) &&
  c0.location=(q2->M1)+(q5->M1)+
      (q1->M2) +(q3->M2)+ (q4->M3) &&
 let c1=c0.next|c1.edges=(q4->q5) && 
 let c2=c1.next|c2.edges=(q3->q5) &&
 let c3=c2.next|c3.edges=(q1->q2 + 
                                  (q2->q3) && 
 let c4=c3.next|c4.edges=(q3->q5) && 
 let c5=c4.next|c5.edges=(q1->q3) && 
 let c6=c5.next|c6.edges=(q2->q4) &&
 let c7=c6.next|c7.edges=(q1->q2)+ 
                    (q2->q4)+(q4->q5) && 
 let c8=c7.next|c8.edges=(q1->q4)+ 
                               (q4->q5) && 
 let c9=c8.next|c9.edges = (q1->q4)+ 
                               (q4->q5) &&
 let c10=c9.next|c10.edges=(q2->q3)+ 
               (q3->q4) + (q4->q5) &&
 let c11=c10.next|c11.edges=(q2 -> q3)+ 
               (q3->q4)+(q4->q5) &&
 let c12=c11.next|c12.edges=(q2->q3)+ 
               (q3->q4)+(q4->q5) &&
 let c13=c12.next|c13.edges=(q->q3)+ 
               (q3->q4) &&
 let c14=c13.next|c14.edges=(q1->q2)+ 
              (q2->q3) &&
 let c15=c14.next|c15.edges=(q1->q5) &&
 let c16=c15.next|c16.edges=(q2->q3)+ 
                  (q3->q4)+(q4->q5) && 
 let c17=c16.next|c17.edges=(q1->q4) && 
 let c18=c17.next|c18.edges=(q2->q4) && 
 let c19=c18.next|c19.edges=(q2->q5) && 
 let c20=c19.next|c20.edges=(q3->q4) }

pred teleport[loc:Qubit->Machine, 
      r:Qubit->Qubit,uloc:Qubit->Machine, 
      tele:Int,utele:Int]
{all disj q0,q1:Qubit|((q0->q1 in r)) implies  q0.uloc=q1.uloc
 utele = plus[tele,#(uloc-loc)] }
fact layerTransition {
 all c:circGraph,uc: grph/next[c] { 
 teleport[c.location,uc.edges,uc.location, 
       c.numTele,uc.numTele] } } 
pred finalLayer {  
	lte[grph/last.numTele,13]  
} 

run finalLayer for 22 circGraph, 7 Int
\end{lstlisting}

\begin{table}
\begin{center}
\begin{tabular}{||c | c | c | c | c||} 
 \hline
 Circuit  & \# of teleported  & Machine 1 & Machine 2 & Machine 3\\ [0.5ex] 
   layer & qubits &  &  & \\ [0.5ex] 
 \hline\hline
 1 & 0 & $q2, q5$ & $q1, q3$  &  $q4$\\ 
 \hline
 2 &  1 & $q2$  & $q1, q3$ & $q4, q5$ \\
 \hline
 3 &  1 & $q2$  & $q1, q3, q5$ & $q4$ \\
 \hline
4 & 1 &  & $q1, q2, q3, q5$ & $q4$ \\
 \hline
 5 & 0 &  & $q1, q2, q3, q5$ & $q4$ \\
 \hline
 6 &  0 &  & $q1, q2, q3, q5$ & $q4$ \\
 \hline
 7 &  1 (swap) &  & $q1, q2, q4, q5$ & $q3$ \\
 \hline
 8 & 0 &  & $q1, q2, q4, q5$ & $q3$ \\
 \hline
  9 & 0 &  & $q1, q2, q4, q5$ & $q3$ \\
 \hline
  10& 0 &  & $q1, q2, q4, q5$ & $q3$ \\
 \hline
   11&  2 &$q1$  & $q2, q3, q4, q5$  &   \\
 \hline
    12&  0 &$q1$  & $q2, q3, q4, q5$  &   \\
 \hline
    13&  0 &$q1$  & $q2, q3, q4, q5$  &   \\
 \hline
    14& 1 (swap) &$q5$  & $q1, q2, q3, q4$  &   \\
 \hline
    15& 0   & $q5$  & $q1, q2, q3, q4$  &   \\
 \hline
    16&  1  & $q1, q5$  & $ q2, q3, q4$  &   \\
 \hline
    17& 1  & $q1$  & $ q2, q3, q4, q5$  &   \\
 \hline
    18& 1  & $q1, q4$  & $ q2, q3, q5$  &   \\
 \hline
    19& 1  & $q1$  & $ q2, q3, q4, q5$  &   \\
 \hline
    20&  0 & $q1$  & $ q2, q3, q4, q5$  &   \\
 \hline
    21&   0 & $q1$  & $ q2, q3, q4, q5$  &   \\
 \hline
    22&  0 & $q1$  & $ q2, q3, q4, q5$  &   \\
 \hline
\end{tabular}
\end{center}
\caption{A solution for the circuit in Figure \ref{fig:circ2}}
\label{tab:sol2}
\end{table}

\section{Minimization and Load Balancing}
\label{sec:loadb}

This section extends the solution proposed in Section \ref{sec:alloySpec} for the cases where teleportations in a DQC network should be performed while ensuring that quantum machines have a balanced load in terms of the number of qubits they hold. Observe that in Table  \ref{tab:sol2} there are rows where Machine 1 and 3 hold no qubits, whereas Machine 2 always holds some number of qubits. {\it What if we want to minimize the number of teleportations while ensuring that the total number of qubits are distributed in a balanced way}? Note that, the requirements of the minimization problem may conflict with the load balancing requirements because distributing qubits evenly may require additional teleportations. Thus, we should optimize two objective functions. To this end, we  first specify the problem of load balancing in the context of our Alloy model. A load balancing scheme should at least result in minimum number of vacant machines in every layer of the circuit as teleportations are minimized. As such, we  minimize two objective functions: the total number of teleportations, and the total number of empty machines. To specify the minimization of vacant machines in our Alloy specification, we need to somehow keep track of such machines in the circuit layers and then ask the Alloy analyzer to minimize the total number of empty machines. Thus, we include an additional relation in the {\mysf circGraph} signature, called  {\mysf emptyMachines} (in Line 5 of Listing \ref{lst:three}). Next, we initialize this relation by including an additional conjunct {\mysf (c0.emptyMachines = 0)} in Line 23 of Listing 2.

\begin{lstlisting}[caption= Relation   {\mysf emptyMachines} captures the total number of vacant machines, language=Alloy,keywordstyle={\bf }, label=lst:three, basicstyle=\small]
sig circGraph{
	edges:Qubit->Qubit, 
	location:Qubit-> Machine, 
	numTele:Int,
	emptyMachines:Int}
\end{lstlisting}

To specify the constraints of both  problems, we revise the {\mysf teleport} predicate as Listing  \ref{lst:four} illustrates. We include two additional input parameters {\mysf emptyMachines} and {\mysf uEmptyMachines} that respectively represent the number of free machines before and after each circuit layer is executed. Line 8-9 of  Listing  \ref{lst:four}  illustrates how we compute  {\mysf uEmptyMachines}. First, notice that,  {\mysf Qubit.uloc} returns all machines that have some qubits located on them. As such, the relation {\mysf Machine - Qubit.uloc} would return all unused machines after {\mysf teleport} is executed on some layer. This value will be added to {\mysf emptyMachines} and assigned to  {\mysf uEmptyMachines} as the new number of vacant machines. To globally minimize the number of unused machines, we include 	{\mysf lte[grph/last.emptyMachines, 19]} in the {\mysf finalLayer} predicate, in addition to already existing constraint {\mysf lte[grph/last.numTele, 13]}. The second parameters of these constraints can initially be some small values and then increased gradually until the Alloy analyzer finds a solution that meets both constraints. Table \ref{tab:sol2} represents the solution generated under load balancing constraints. Without these constraints, we get a solution in which the machine M3 is never used!

\begin{lstlisting}[caption= Revised {\mysf teleport} predicate models how vacant machines are counted, language=Alloy,keywordstyle={\bf }, label=lst:four, basicstyle=\small]
pred teleport[loc:Qubit->Machine, 
  r:Qubit->Qubit,uloc:Qubit->Machine,
  tele:Int,utele:Int,
  emptyMachines:Int,uEmptyMachines:Int]
 {all disj q0,q1:Qubit|
   (q0->q1 in r) implies q0.uloc=q1.uloc
 utele=plus[tele,#(uloc-loc)]
 uEmptyMachines= 
  plus[emptyMachines,#(Machine-Qubit.uloc)]}
\end{lstlisting}

\section{Minimization in Heterogeneous Networks}
\label{sec:hetero}

This section investigates the problem of minimizing the number of teleportations in heterogenous quantum networks where the cost of teleportations are non-uniform from one machine to another. Thus far, we have assumed that teleporting a qubit from one machine to any other machine has a unit cost. This is an unrealistic assumption in part because the teleportation costs differ depending on the geographical distance of quantum machines from each other. To capture this reality, we include the new relation {\mysf costTo} in the {\mysf Machine} signature in Line 2 of Listing \ref{lst:five}. The ternary relation {\mysf Machine $\rightarrow$ Machine $\rightarrow$ Int} associates an integer weight to each network link in order to  represent the cost of teleporting from one machine to another.  The fact in Lines 3-8 in Listing \ref{lst:five} hard codes the relation {\mysf costTo} by assigning weight to network links. Lines 7 and 8 stipulate that teleporting from a machine to itself costs nothing. Moreover, the  {\mysf circGraph} signature includes an additional relation {\mysf teleCost} in order to model the total cost of teleportations. The logic behind specifying   {\mysf teleCost} is similar to our reasoning for the inclusion of  {\mysf emptyMachines} as described in the previous section. 

\begin{lstlisting}[caption= Modeling the heterogeneous cost of teleportation, language=Alloy,keywordstyle={\bf }, label=lst:five,basicstyle=\small]
abstract sig Machine {
	costTo:Machine->Int } 
fact {
 costTo =  (M1 -> M2 ->1)+(M1 -> M3 ->2)+
  (M2 -> M1->1)+(M2 -> M3 ->1)+
 (M3 -> M1->2)+(M3 -> M2 ->3)+
 (M1 -> M1->0)+(M2 -> M2 ->0)+  
 (M3 -> M3 ->0)  }
sig circGraph{
	edges: Qubit->Qubit, 
	location: Qubit-> Machine, 
	numTele:Int, 
	emptyMachines:Int, 
  // Models the cost of teleportations
	teleCost:Int } 
pred teleport[loc:Qubit->Machine,
  r:Qubit->Qubit,uloc:Qubit->Machine,
  tele:Int,utele:Int,emptyMachines:Int,
  uEmptyMachines:Int, 
  totCost:Int,uTotCost:Int]
  {all disj q0,q:Qubit| 
         (q0->q1 in r) implies q0.uloc=q1.uloc
  utele = plus[tele,#(uloc-loc)]
  uEmptyMachines = 
    plus[emptyMachines,#(Machine-Qubit.uloc)]
  uTotCost = 
    plus[totCost, 
       sum q:Qubit|((q.loc).costTo)[q.uloc]] }
fact layerTransition {
 all s:circGraph,us:grph/next[s] { 
    teleport[us.edges,s.location,us.location, 
        s.numTele,us.numTele, 
        s.emptyMachines,us.emptyMachines,
        s.teleCost,us.teleCost] }}
pred finalLayer {  
	lte[grph/last.numTele,13]  
	lte[grph/last.emptyMachines,19]  
	lte[grph/last.teleCost,11]  } 
\end{lstlisting}
 
 Listing \ref{lst:five} presents the static part of the changes that we make in our Alloy specification in order to generalize the problem for heterogeneous networks. Nonetheless, we still need to revise the predicates {\mysf teleport} and  {\mysf finalLayer} towards encoding the dynamic aspects of our model. To this end, we first initialize the  {\mysf teleCost} relation by including an additional constraint {\mysf (c0.teleCost = 0) } to Line 23 of Listing \ref{lst:two}. To add up the heterogeneous cost of teleportations, we include the constraint {\mysf uTotCost} in Lines 26-28 of Listing \ref{lst:five}, where the quantifier {\mysf sum} is used. This quantifier acts akin to a universal quantifier except that the expression `{\mysf ((q.loc).costTo)[q.loc]}' is evaluated for each quantified qubit $q$, and the evaluated values are summed up and returned. The constraint `{\mysf ((q.loc).costTo)[q.loc]}' simply returns the cost of teleporting a qubit $q$ from its current location to another machine in the network. The {\mysf layerTransition} fact is revised accordingly (in Line 34). Finally, the additional predicate {\mysf lte[grph/last.teleCost ,11]} is included inside the predicate {\mysf finalLayer} in Line 38 of Listing \ref{lst:five} whose logic is similar to those of Lines 36 and 37. In fact, {\mysf finalLayer} minimizes three objective functions simultaneously: the TMP problem (Line 36), load balancing (Line 37) and the communication costs of the heterogeneous network (Line 38).

\section{qcAlloy Architecture and Experimental Evaluation}
\label{sec:eval}

This section presents the details of  qcAlloy, which is a realization of the framework of Figure \ref{fig:vision} in the context of Alloy. We also present the results of our experimental evaluations while comparing them with a state-of-the-art method. Section \ref{machine-format} describes the format of the input file to qcAlloy. Then, Section \ref{qcAlloyArch} presents the architecture of qcAlloy, and finally, Section \ref{exprRes} discusses our experimental results.

\subsection{Machine-Readable Circuit Format}
\label{machine-format}

qcAlloy supports the machine-readable .tfc file format, which is used for the textual specification of a large quantity of circuits described in \cite{revLib2005}.
An example 3-bit adder circuit in the .tfc format is shown in Listing \ref{lst:tfc}.
The format begins with a header containing metadata, including qubit variable lists for all variables (.v), input variables (.i), output variables (.o), and constant variables (.c).
The body of the format contains a sequence of gates between a BEGIN keyword and a concluding END keyword.
Each gate is described as a list of the qubits involved prefaced with a 't' for Toffoli gate or an 'f' for Fredkin gate along with the gate arity.
For our purposes, only the metadata list of all variables and the variable list for each gate are required.

\begin{figure}[htbp]
\captionof{lstlisting}{Example .tfc file describing a 3-bit adder from \cite{revLib2005}}
\label{lst:tfc}
\begin{minipage}{0.5\columnwidth}
\begin{lstlisting}[basicstyle=\small]
  .v a,b,c,d
  .i a,b,c
  .o d,c
  .c 0
  BEGIN
\end{lstlisting}
\end{minipage}%
\begin{minipage}{0.5\columnwidth}
\begin{lstlisting}[firstnumber=6, basicstyle=\small]
  t3 a,b,d
  t2 a,b
  t3 b,c,d
  t2 b,c
  END
\end{lstlisting}
\end{minipage}
\end{figure}

\subsection{The Architecture of qcAlloy}
\label{qcAlloyArch}
\begin{figure}
    \centering
    \includegraphics[scale=0.34]{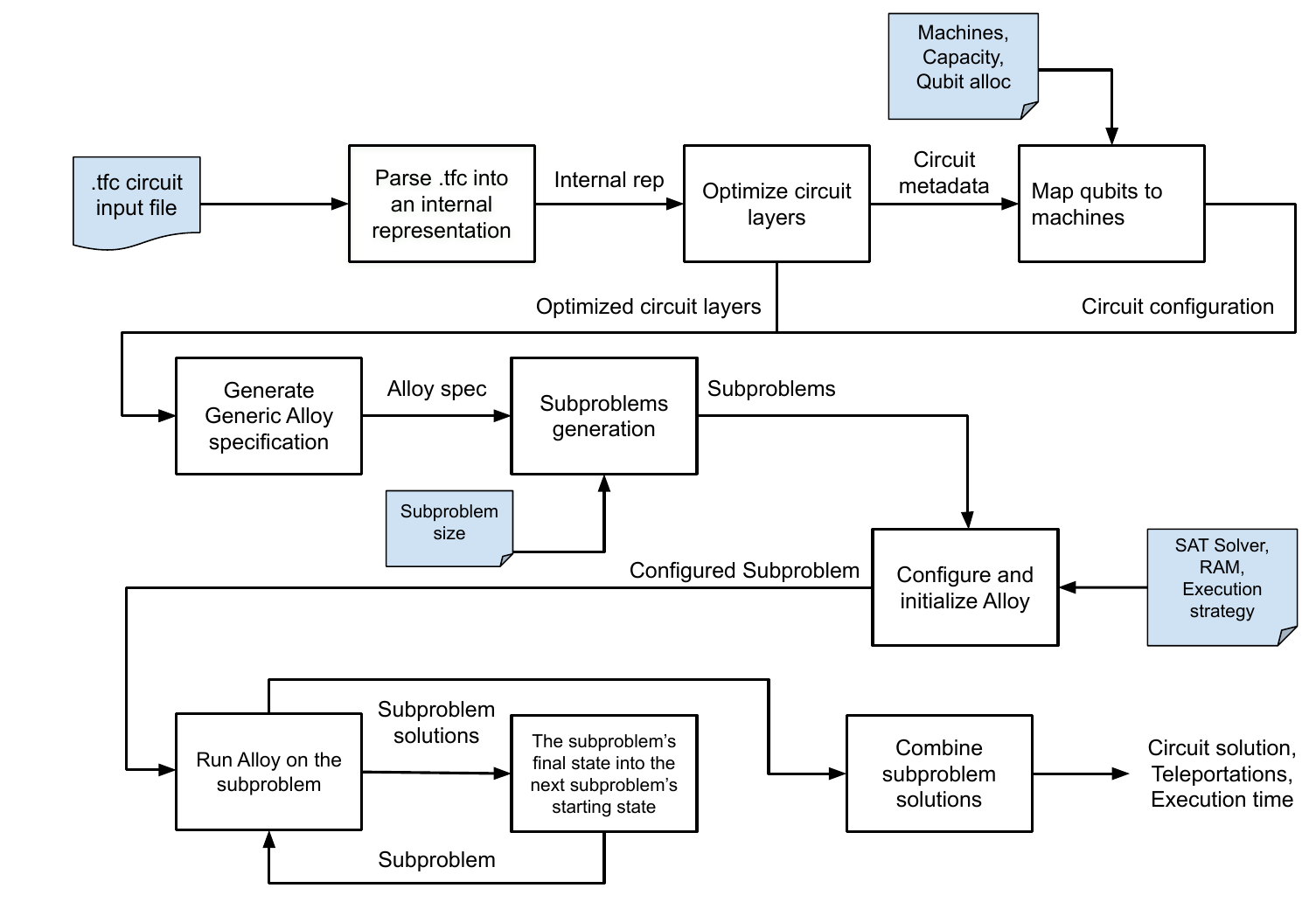}
    \caption{The architecture of qcAlloy}
    \label{fig:qcAlloy}
\end{figure}

The qcAlloy tool takes a textual file (in .tfc format) and  generates the corresponding Alloy model. The Alloy model is then run by the Alloy analyzer, which transforms the Alloy model into a Boolean satisfiability instance (i.e., a CNF formula) and invokes one of several integrated SAT solvers. If the SAT instance is satisfiable, a solution is generated that describes a scenario of how qubits should be teleported  towards solving the TMP (stated in Section \ref{sec:prob}).
Figure \ref{fig:qcAlloy} illustrates the architecture and the workflow of qcAlloy.

\noindent{\bf Input Circuit Description}.
Starting with the .tfc file describing the circuit, qcAlloy parses the file into an internal representation which is used to optimize the circuit.
The first optimization drops gates with an arity of one as such gates do not require teleportation and cannot affect a solution.
The second optimization identifies and organizes the circuit layers. 
Both of these optimizations decrease the number of layers needed to describe the circuit.

\noindent{\bf Problem Parameters}.
After  the circuit is analyzed, qcAlloy takes some parameters to create a complete formal specification of the problem in Alloy, called the Alloy model. 
These parameters include the number of machines, the qubit capacity of the machines, and the qubit allocation policy. In our current experiments, we have considered a fully connected network of quantum machines. 
The initial qubit allocation policies explored in this work are `random' and `in-order'.
The random policy distributes qubits to machines randomly while the in-order policy assigns qubits by filling up the machines in-order and moving onto the next machine once the capacity of the current machine is reached.
Based on the allocation policy, the qubit variables of the circuit are assigned to the quantum machines.
This assignment will be the starting state for the Alloy model.
The reusable and extensible Alloy model can then be completed by including the circuit structure and the starting qubit-to-machine mapping.
Any other additional constraints can be manually included in the Alloy model at this stage.

\noindent{\bf Subproblem Generation}.
The next step is the generation of subproblems.
As Alloy models can become prohibitively expensive as the circuit size increase, qcAlloy has the option of splitting up the input circuit.
A subproblem size, in terms of the maximum number of layers for each subproblem, is given as input to qcAlloy.
The circuit can then be split into {\it circuit\_layers/subproblem\_layers} number of subproblems.
The final qubit-to-machine mapping from the execution of a given subproblem becomes the starting state of the next.

\noindent{\bf Alloy Configuration}.
After  subproblem generation, we configure the SAT solver integrated in Alloy.
Configuration options include which SAT solver to use, the RAM available to Alloy, and the execution strategy.
In this work, we use the MiniSATJNI SAT solver with a modified binary search execution strategy.

\noindent{\bf Model Execution Strategy}.
The execution strategy describes how Alloy determines the solution to each subproblem.
Since Alloy gives a `satisfiable' or `unsatisfiable' response, we specify the Alloy model as a decision problem by including a threshold parameter, which captures an upper bound for the number of allowable teleportations. Then, we start from a small threshold where the Alloy model is `unsatisfiable' and gradually increase the threshold until we reach the first value of the threshold parameter for which the model is `satisfiable' (indicating the minimum number of teleportations for which Alloy can find a solution). We denote this method as the \textit{linear} execution strategy. To improve the efficiency of this process for circuits with a high number of teleportations, we devise the binary search execution strategy as an alternative. 
The binary search algorithm reduces the complexity of executing a subproblem  from $O(n)$ to $O(log(n))$, where $n$ denotes the maximum number of teleportations as a multiple of the number of layers. 
However, we found that this can be improved  further in practice. In this work, we use a modified binary search algorithm that keeps a history of the previous  results from the last, say ten, subproblems.
The minimum and maximum values of this history are used as the starting range for subsequent binary searches.
If this range fails to find the minimum number of teleporations, the range is expanded to $n$.
We found experimentally that solutions to previous subproblems often accurately inform later subproblems. Thus, this modified approach reduces execution time costs further, especially when there are many subproblems from which a history can be exploited.

\noindent{\bf Running the Model}.
The Alloy model executes starting with the first subproblem and in accordance with the execution strategy.
The next subproblem is given the previous subproblem's solution to be used as the starting state.
The current subproblem is then executed in accordance with the execution strategy until a solution is found.
This process continues until all subproblems have been executed.

\noindent{\bf Combining the solutions of subproblems}.
After having executed all subproblems in sequence, the solution to each subproblem is combined into a single solution.
An analysis is also performed to determine which teleportations were swap operations.

\subsection{Results and Comparative Analysis}
\label{exprRes}
\begin{figure}
\centering
\includegraphics[scale=0.12]{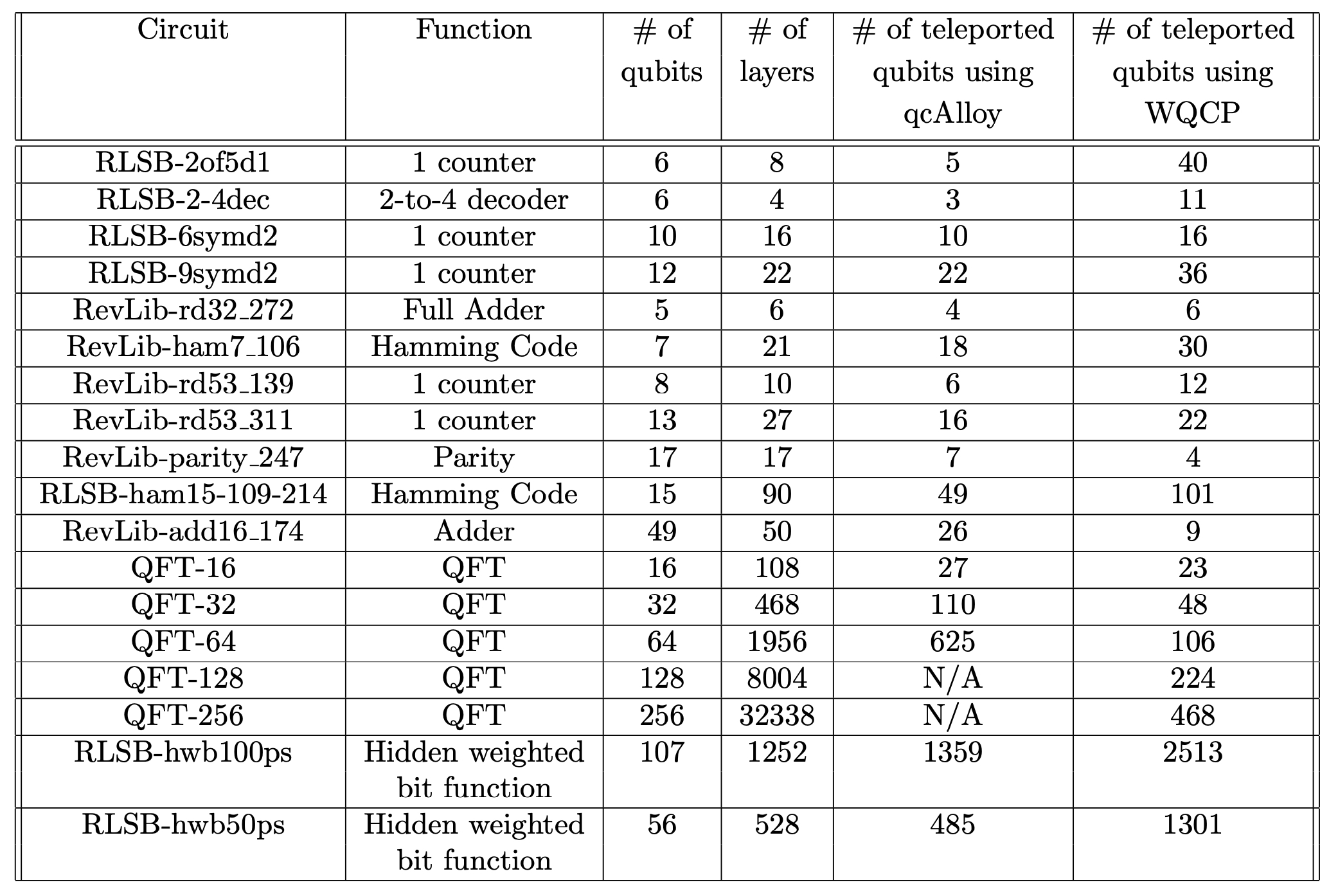}
\caption{Number of teleportation  in qcAlloy vs. the Window-based Quantum Circuit Partitioning (WQCP) approach in \cite{nikahd2021automated}.} 
\label{fig:eval}
\end{figure}

We evaluate qcAlloy using circuits from the Reversible Logic Synthesis Benchmarks Page (RLSB) \cite{revLib2005} and the RevLib benchmark database \cite{wille2008revlib} which are in the .tfc machine-readable format (see Section \ref{machine-format}).
We also evaluate qcAlloy on  the Quantum Fourier Transform (QFT) circuit from 16 to 256 qubits. 
The results are given in Figure \ref{fig:eval}.
Any circuits obtained from one of the libraries are prefixed with either RLSB or RevLib respectively.
All results were gathered from a dedicated machine with an i7-12700 CPU and 32GB of RAM. Benchmarks were run in parallel with no more than four at any given time such that execution times remain independent.

The table in Figure \ref{fig:eval} presents the results of the benchmarks run using qcAlloy and the results from the Window-based Quantum Circuit Partitioning (WQCP) approach \cite{nikahd2021automated} as the baseline of this work.
The number of teleportations reported by each approach are given along with the circuit function and the number of qubits and the number of layers after optimization in the circuit.
The WQCP approach runs each benchmark multiple times across a range of configurations of their algorithm.
We show the result of the configuration with the minimum number of teleportations for each benchmark. 

The qcAlloy approach provides a lower number of required teleportations than WQCP for 11 out of the 18 benchmarks evaluated while remaining competitive in the remainder. 
As smaller circuits do not need to be split into multiple subproblems, the model can search across the entire circuit, leading to more optimal solutions.
The larger circuits, however, need to be split into subproblems. 
Higher teleportations may be reported because finding the solution of a subproblem is currently done without considering the nature of the next subproblem. 
For example, the model may perform a teleportation swap without considering the fact that such a teleportation will make the solution to the next subproblem non-optimal. 
 QFT provides an exceptional challenge for qcAlloy. We believe that QFT requires a different strategy for subproblem generation as QFT's structure follows a recursive pattern.
However, the hwb100ps and hwb50ps circuits provide a solution about half that of WQCP. 
We conjecture this is the case due to the seemingly random collection of gates and qubits.
Such a distribution does not punish short-term decisions by the Alloy model which may also appear random holistically.

\section{Related Work}
\label{sec:rel}

This section discusses the most important work related to the proposed approach. In general, we observe three  families of methods in the literature, namely {\it graph-theoretic approaches, custom-designed heuristics} and {\it DQC compilers}. As an example of graph-theoretic techniques, Andres-Martinez and Heunen \cite{andres2019automated}  reduce the minimization problem to the problem of hypergraph partitioning where   the number of cuts in the partitioned graph must be minimized. First, they transform the quantum circuit to a hypergraph whose vertices include qubits and gates and each hyperedge connects a subset of vertices. Then, they conduct a multi-step preprocessing phase where they transform the circuit into a binary circuit, reorder CZ gates and move them ahead so they can be grouped  (may introduce some single-qubit gates),  and then split the circuit into multiple segments. While this approach is efficient for some circuits, the hypergraph partitioning is a bottleneck because it is in general NP-hard. Davarzani  {\it et al.}   \cite{davarzani2020dynamic} create a bipartitie graph out of a quantum circuit where the two sets of vertices include the qubits and the gates. Then, they present a dynamic programming algorithm for partitioning the set of qubits and calculating the minimum number of teleportations for each way of partitioning. Their algorithm uses memoization to avoid recalculating repeated sub-problems. Daei {\it et al.}   \cite{daei2020optimized} model a quantum circuit as a weighted undirected graph where vertices represent the qubits, edges capture the input qubits of binary and ternary gates, and edge weights represent the number of gates that use each pair of qubits. Then, they apply the graph partitioning method of Kernighan–Lin \cite{kernighan1970efficient} to find the partition that has the minimum-weight cut. 

Heuristic-based approaches develop special heuristics to improve the efficiency of minimization  in terms of both execution time and the number of teleportations. For example, Nikahd  {\it et al.}    \cite{nikahd2021automated} present a window-based partitioning method where they combine gate and qubit partitioning. First, they  identify the layers of a circuit as sub-circuits that contain concurrent gates. Second,  starting from the first layer (i.e., the leftmost layer), they consider a window of length $w$ that slides from left to right and creates sub-circuits. Third, they formulate the qubit partitioning of the sub-circuits as an Integer Linear Program (ILP), which they solve using the CPLEX ILP solver.  Ranjani and Gupta \cite{sundaram2023distributing} present two algorithms: a local-best algorithms and a zero-stitching algorithm. Their local-best algorithm uses Tabu search to initially map qubits to machines in such a way that the number of global gates is initially minimized. Then, they scan the circuit from left to right, and for each global gate $g$ they determine a machine $p$ in which $g$ should be executed. Based on this, they teleport the qubits to $p$. To select a machine $p$, they define a benefit function for $p$ based on the number of global gates that become local in the next $r$ levels of the circuit if $p$ is used for executing $g$. In the zero-stitching method, they use a dynamic programming approach to partition the circuit into sub-circuits that can be executed without any teleportations, and then stitch them together. Nonetheless, teleportations may be performed at the stitching points. Zomorodi-Moghadam {\it et al.}   \cite{zomorodi2018optimizing} consider quantum circuits for a distributed quantum network of only two machines. Their algorithm explores all possible configurations of executing every global gate in either one of the machines, and finds the  number of teleportations for each configuration. Then, they take the minimum of all possible teleportation scenarios.

Most recent DQC compilers focus on identifying communication patterns in quantum circuit as well as implementing  minimized teleportation plans on quantum networks. For example, Wu  {\it et al.}  \cite{wu2022autocomm} observe that many remote  two-qubit gates can be executed using one or two quantum communications. To this end, they define the notion of {\it burst communication} that captures a group of continuous global two-qubit gates between one qubit and one node. Their objective is to maximize the number of global gates that can be executed using each entangled pair. They develop a compiler centered around the concept of burst communication, where a quantum circuit is initially processed to identify burst communication blocks. 
 In another work, Wu  {\it et al.}  \cite{wu2023qucomm} present a compiler for mapping a circuit to a DQC network while minimizing the communication costs. Their approach includes three major phases: (1) identify collective communication blocks, where each block is a set of global gates whose pattern of qubit communication forms a connected graph over multiple network nodes; (2) perform optimal data routing in each communication block under the topological constraints of the network, and (3) use data qubits as a buffer for communication qubits (i.e., qubits used in entangled pairs). They find data qubits that incur the minimum communication overhead. 
 Ferrari {\it et al.}  \cite{ferrari2021compiler}  discuss the challenges of developing compilers for DQC, and then present an upper bound complexity/overhead for the compilation process. They point to the remote operations as the main challenge, and state that two types of constraints govern the costs of remote operations: the topological constraints of the underlying network, and the limitations in having and regenerating entangled pairs.

While the aforementioned approaches solve the minimization problem in many cases, most of them can  process only circuits with binary or ternary gates, and provide little reuse of problem specifications (e.g., \cite{andres2019automated}). By contrast, the proposed approach in this paper can solve the problem for general quantum circuits with $n$-ary gates. Moreover, the Alloy specification of the problem is reusable and can easily be instantiated for different circuits/networks. The Alloy specifications can also  be easily extended to address other problems such as load balancing and minimization in heterogeneous quantum networks (as demonstrated in Sections \ref{sec:loadb} and \ref{sec:hetero}).

\section{Conclusions and Future Work}
\label{sec:concl}

This paper presented a novel approach for solving the Teleportation Minimization Problem (TMP) using the formal specification language Alloy \cite{jackson2012software} and its analyzer. The proposed framework (in Figure \ref{fig:vision}) enables a transformative method for enabling reuse and flexibility in tackling the TMP problem in particular, and solving the distribution of quantum algorithms, in general. The reusable Alloy specifications remain unchanged when moving from one circuit/network to another, thereby providing tremendous simplicity and transparency for mainstream engineers when it comes to specifying the general constraints of mapping quantum algorithms to networks. The circuit/network-dependent part of the formal specifications are automatically generated, and combined with the reusable part. The Alloy analyzer then searches for a scenario of qubit routing that incurs minimum number of teleportations. The implementation of the proposed approach as a software tool, called qcAlloy, is available at \url{https://github.com/KieranYoung/Alloy-DQC}. Our experimental results in the context of the RevLib benchmark \cite{revLib2005,wille2008revlib} indicate that qcAlloy can solve the TMP problem for circuits with more than 100 qubits and 1200 layers. qcAlloy outperforms one of the most efficient methods \cite{nikahd2021automated}  in terms of the number of teleportations in most cases up to 50\% while underperforming in a few cases (e.g., QFT circuit). 

There are some {\bf limitations} that lay the ground for our ongoing and future work. First, the Alloy execution time in some cases becomes prohibitive in part due to the inefficiency of the selected SAT solver for that particular circuit/network. As such, determining which SAT solvers perform better remains an open problem. Second, in the execution strategies, subproblems are run multiple times, which incurs a high execution time. We plan to address this by developing alternative strategies. For instance, estimations of the minimum teleportation solution may be sufficient and drastically save on execution time. We also look to pair this with increasing accuracy by supplying subproblems with a lookahead mechanism so that unnecessary teleportations are prevented. Third, identifying recurring architectural patterns in the input quantum circuit can help in the reuse of solutions. Such recurring patterns can also help in devising more intelligent divide and conquer methods for circuit partitioning. Fourth, in  large circuits,  the gate arity is often well below the qubit count. However, in our current Alloy specifications we still include the unused qubits in a layer despite the fact that many of these qubits may not appear in the current subproblem. As a result, the scope of integers during Alloy verification is increased, which has a direct negative impact on execution  time of the Alloy analyzer. Thus, an optimization is to strip the circuit of qubits that are seldom used globally or for the current subproblem. This could also be done for some number of machines that are deemed unnecessary to be considered for the current subproblem. Fifth, we plan to develop a repository of formal specifications of the problems that should be solved in the compilation of quantum circuits for DQC. Such problems include the TMP for logical qubits under a variety of network constraints (e.g., topology, communication costs), the routing of logical qubits under network constraints, qubit buffering, etc. Such a repository will be a valuable asset for researchers as everyone can reuse them, and new specifications can be added for new problem variants. The entire collection will form a public library of formal specifications for DQC compilers. Moreover, the correctness of different compiler designs can be verified against formal specifications, thereby providing high assurances regarding the compilation process. Finally, we would also like to investigate the effectiveness of our approach in the transpilation of quantum circuits on noisy quantum machines.

\bibliographystyle{IEEEtran}
\bibliography{biblio}

\end{document}